\documentclass[runningheads]{llncs}

\usepackage{color,soul}
\usepackage{cite}
\usepackage{amsmath,amssymb,amsfonts}
\usepackage{algorithmic}
\usepackage{graphicx}
\usepackage{textcomp}
\usepackage{xcolor}
\usepackage{framed,multirow}
\usepackage{mathtools}
\usepackage{tabularx,booktabs}
\usepackage{gensymb}

\DeclarePairedDelimiter\floor{\lfloor}{\rfloor}
\newcolumntype{Y}{>{\centering\arraybackslash}X}

\begin{document}
\title{Interpreting Age Effects of Human Fetal Brain from Spontaneous fMRI using Deep 3D Convolutional Neural Networks}
%
%


\author{Xiangrui Li\inst{1} \and
		Jasmine Hect\inst{2}\and
		Moriah Thomason\inst{3}\and
	    Dongxiao Zhu\inst{1}
}

\authorrunning{Xiangrui et al.}
%
\institute{ Department of Computer Science, Wayne State University \email{\{fx7219,dzhu\}@wayne.edu} \\
		\and
	Department of Psychology, Wayne State University \\ \email{ev4125@wayne.edu}\\
	\and 
	Department of Child and Adolescent Psychiatry, New York University \email{moriah.thomason@nyulangone.org}
}
\maketitle              
\begin{abstract}
Understanding human fetal neurodevelopment is of great clinical importance as abnormal development is linked to adverse neuropsychiatric outcomes after birth. Recent advances in functional Magnetic Resonance Imaging (fMRI) have provided new insight into development of the human brain before birth, but these studies have predominately focused on brain functional connectivity (i.e. Fisher z-score), which requires manual processing steps for feature extraction from fMRI images. Deep learning approaches (i.e., Convolutional Neural Networks (CNN)) have achieved remarkable success on learning directly from image data, yet haven’t been applied on fetal fMRI for understanding fetal neurodevelopment. Here, we bridge this gap by applying a novel application of deep 3D CNN to fetal blood oxygen-level dependence (BOLD) resting-state fMRI data. Specifically, we test a supervised CNN framework as data-driven approach to isolate variation in fMRI signals that relate to younger \textit{v.s.} older fetal age groups. Based on the learned CNN, we further perform sensitivity analysis to identify brain regions in which changes in BOLD signal are strongly associated with fetal brain age. The findings demonstrate that deep CNNs are a promising approach for identifying spontaneous functional patterns in fetal brain activity that discriminate age groups. Further, we discovered that regions that most strongly differentiate groups are largely bilateral, share similar distribution in older and younger age groups, and are areas of heightened metabolic activity in early human development.

\end{abstract}

\section{Introduction} \label{sec:intro}
The human brain undergoes rapid development in the prenatal period. Previous studies show significant links between abnormal brain development \textit{in utero} and adverse postnatal outcomes \cite{benkarim2017toward,van2016functional}. As a result, understanding fetal neurodevelopment, particularly timescales and patterns for maturation of functional systems, is of great clinical importance. 

Recent methodological advances in fetal functional Magnetic Resonance Imaging (fMRI) have attracted much attention \cite{konkel2018brain,van2016functional}. One promising advance is application of CNN-based methods to fetal brain MRI image segmentation \cite{makropoulos2018fetalsegmentation}. For fetal neurodevelopmental studies, existing works focus on analysis of the functional connectivity in fetal brain, such as discoveries of functional networks and identification of highly connected hubs \cite{van2018hubs,thomason2015age}. Those studies are based on functional connectivity measures (i.e. adjacency matrix of Fisher z-scores) extracted from the spatio-temporal 4D resting-state fMRI. However, from the perspective of computational methodology, those approaches rely heavily on the data processing steps for extracting z-scores and only utilize temporal information in the fMRI images, which miss the spatial information in the 2D or 3D fMRI images.

There is precedent for direct learning from MRI images in children and adults. For example, CNN has been used to classify neurological disorders such as Autism Spectrum Disorder (ASD) \cite{li2018asd,zhao2018asd} and Alzheimer's disease \cite{hosseini2016alzheimer,korolev2017alzheimer}, and to characterize brain functional networks \cite{zhao2018automatic}. Further, a small number of studies have built CNN models to predict age from fMRI data \cite{li2018brainage}. Probably due to very limited availability of fetal fMRI data, CNN models have not been applied for understanding fetal neurodevelopment.

To bridge the methodological gap, we propose a novel application of deep CNN in conjunction with CNN interpretation techniques on the analysis of fetal brain fMRI data, by leveraging CNN's merits of learning high-level feature representations. To the best of our knowledge, this work represents the first application of CNN models as a data-driven approach in understanding fetal developmental processes. Specifically, as age and brain development are closely linked, we utilize the age group as the surrogate target (younger \textit{v.s.} older) and train a supervised CNN directly on the residual fMRI images to capture associations between age effect and blood oxygen-level dependence (BOLD). The trained model achieves superior predictive performance over traditional machine learning algorithms, demonstrating its effectiveness in learning directly from fetal fMRI images. More importantly, since CNN is interpretable, we are able to explain the trained CNN to highlight brain regions that are strongly influential in model predictions. The interpretation results of important regions are clinically compelling that match with existing studies. Furthermore, the trained CNN model also proposes novel clinically sensible regions that are potentially informative in characterizing brain development in fetuses, demonstrating its strong promise as a data-driven approach to enhance our understanding of brain development as well as facilitate fetal developmental studies.

\section{Method}

\begin{figure}[t]
	\centering
	\includegraphics[scale=0.25]{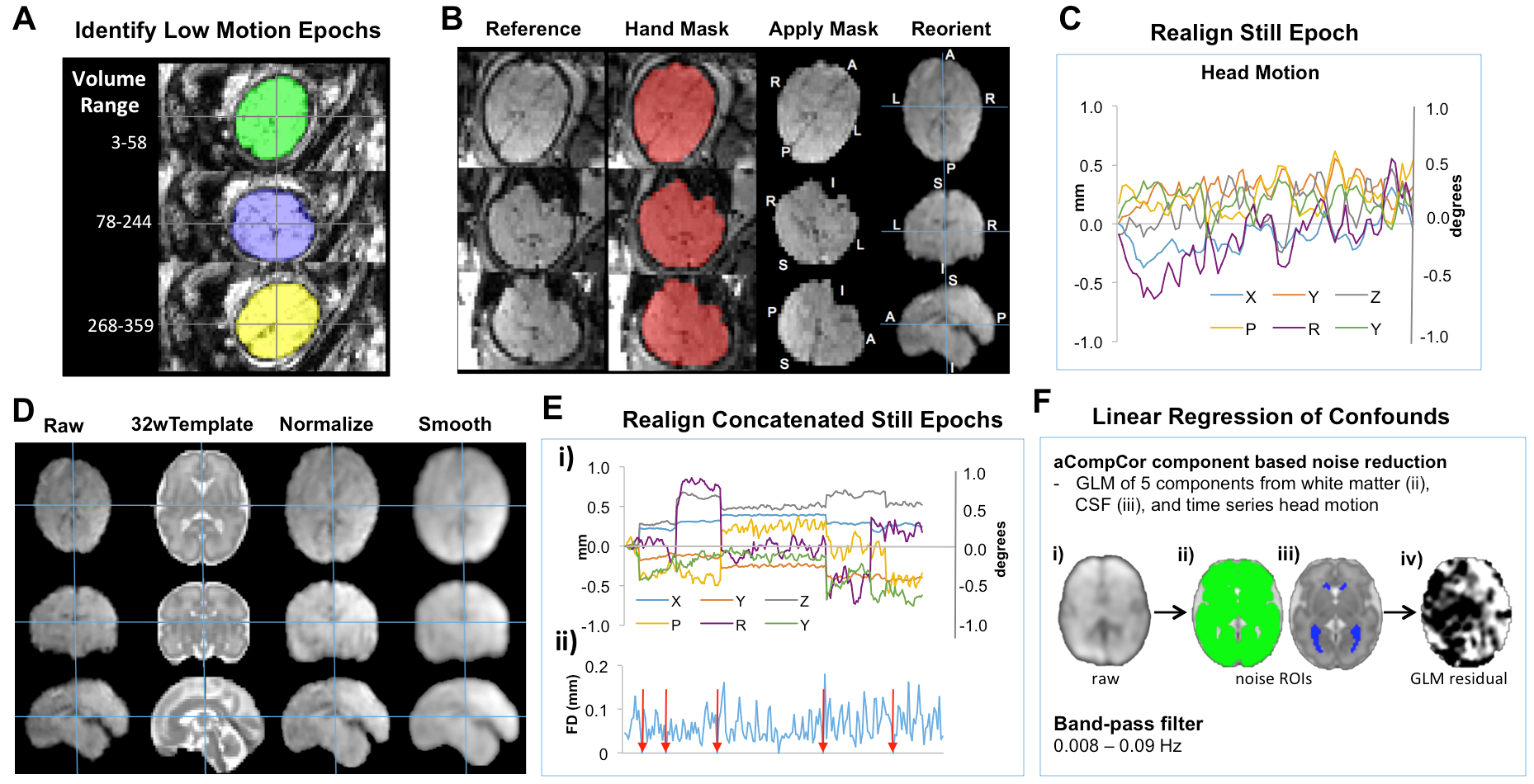}
	\caption{Data preprocessing pipeline from raw MRI image to residual fMRI data.}
	\label{fig:imgPreprocessing}
\end{figure}
\subsection{Data Acquisition and Preprocessing}

Subjects in our analysis were subset of a cohort in a study of longitudinal fetal brain development. The cohort consists of 148 pregnant women that underwent MRI with gestational age (GA) between 24 and 40 weeks. We manually checked subjects according to data quality (high motion, artifacts et al.) and inclusionary criteria  (gestational age, born without nervous system abnormality et al.), leaving 75 qualified subjects: 30 subjects with GA between 26 and 29 weeks (younger group), and 45 between 34 and 37 weeks (older group).

MRI data were acquired on a Siemens Verio 70-cm open-bore 3T MR system. Resting-state fMRI data were acquired using a gradient echo planar imaging sequence: TR/TE 2000/30 ms, flip angle $80^\circ$, 360 frames, axial 4 mm slice thickness, voxel size 3.4x3.4x4 mm$^{3}$, repeated twice. Between 12 to 24 minutes of fetal resting-state fMRI data were collected per subject.

Extraction of fetal brain fMRI data requires multiple steps in preprocessing raw resting state fMRI images. In brief, periods of fetal quiescence were manually identified using FSL image viewer, wherein individual segments must consist of at least 20 seconds (10 frames) of low motion ($<$2 mm translation and/or 3 degrees rotational movement) (A). After motion censoring, fetal brain masks were then created separately for each low-motion epoch from a single reference image using Brainsuite \cite{shattuck2002brainsuite} (B). After masking, each temporal segment was reoriented manually, realigned to the mean BOLD volume, resampled to 2 mm isotropic voxels, and normalized to a 32-week fetal brain template \cite{serag2012temp} using SPM8 \cite{penny2011spm} (C). All normalized images from each segment were then concatenated into one run, realigned to the mean BOLD volume, and smoothed with a 4 mm FWHM Gaussian kernel (D-E). Further preprocessing was performed in CONN toolbox (v14n) \cite{whitfield2012conn} including linear detrending, nuisance regression using aCompCor of five principal components extracted from a 32-week fetal atlas white matter and CSF mask, six head motion parameters, and band-pass filtering at 0.008 to 0.09 Hz (F). Fig. \ref{fig:imgPreprocessing} shows the preprocessing pipeline from raw MRI to residual fMRI used in our model.

\subsection{Modeling Age Effect on Neurodevelopment with 3D CNN}
For each subject, the data are a time series of 3D fMRI volumes of dimension $43\times 51 \times 40$. To exploit the spatio-temporal information in fMRI, we take a sliding window approach \cite{li2018asd}. Specifically, starting from the first 3D frame of 4D fMRI whose length is $T$, we take the mean 3D fMRI image in a window of size $m$ over the time axis; as the window slides over the time axis with stride $s$, a total of $\floor*{\frac{T-m}{s}}+1$ 3D fMRI images are generated, which significantly increases the sample size in training data. After initial experiments, $m=2$ (or $3$) and $s=1$ achieve good performance and increasing window size leads to over-smoothing and degrade model performance. As mentioned in Section \ref{sec:intro}, younger (26$<$GA$<$29 weeks) and older (34$<$GA$<$37 weeks) age groups are used as the surrogate target. For each 3D fMRI generated from sliding window, we label it with the subject's age group. 

\begin{figure}[t]
	\centering
	\includegraphics[width=\columnwidth]{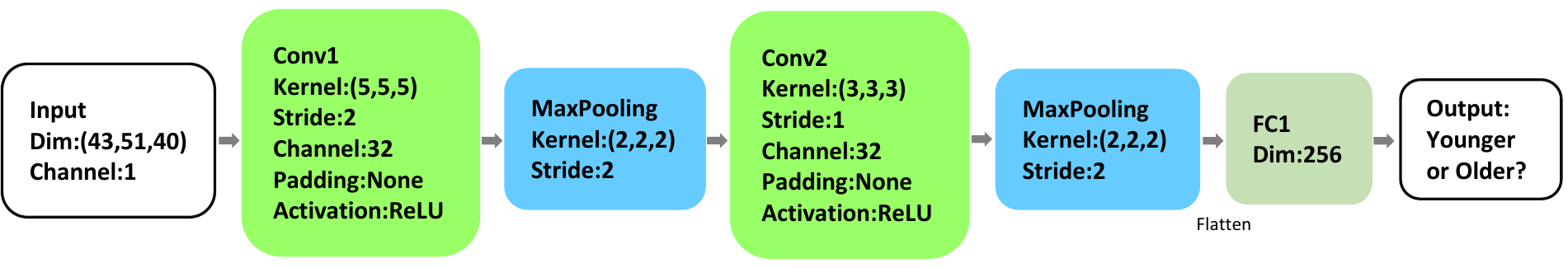}
	\caption{CNN architecture for age group prediction.}
	\label{fig:cnnArch}
\end{figure}

To learn fetal neurodevelopmental processes by classifying 3D fMRI, we build an effective 3D CNN to capture the association between residual fMRI and age group. The proposed CNN along with its architectural parameters (i.e. kernel size, stride, channel et al.) is shown in Fig. \ref{fig:cnnArch}. It has two convolutional layers with ReLU non-linear activation and one fully connected layer, where each convolutional layer is followed by a max pooling layer. Model predictions are made with sigmoid activation on the output layer after the fully connected layer. The objective function is hence the binary cross entropy plus $L_2$ regularization on the network weights:
$$ L(\boldsymbol{\theta}) = -\sum_{i} y_i \log(f(\boldsymbol{x}_i)) + (1-y_i) \log (1-f(\boldsymbol{x}_i)) + \lambda ||\boldsymbol{\theta}||_2^2,$$
where $y_i=1$ (older group) or $0$ (younger group) is the true label of $\boldsymbol{x}_i$, $f(\boldsymbol{x}_i)$ is the probability $p(y=1|\boldsymbol{x}_i)$ modeled by CNN, and $\lambda$ is the tuning parameter of $L_2$ regularization.\smallskip

\subsection{Model Interpretation} \label{sec:modelIntpretation}
As our primary goal is to understand the age effect on fetal brain development, model interpretation is critical to our CNN approach. Once CNN is trained, we perform sensitivity analysis (SA) \cite{simonyan2013sa,rieke2018vis} on leave-out testing images to identify important regions of interests (ROIs). The sensitivity score for each image pixel is the squared gradient with respect to the CNN output (i.e. probability of being younger or older), which measures how sensitive the prediction is to the change of voxel values. The calculation of sensitivity score only needs one pass of backpropagation with respect to the input image. For fetal fMRI images, a region (i.e. multiple voxels in a neighborhood) in the brain with larger sensitivity scores is more influential, hence more important in predicting age group.

\subsection{Training Setup}
The CNN is trained with classic stochastic gradient decent algorithm (SGD) with momentum set to 0.8. The learning rate is set to 0.1 initially and decreased by multiplying a factor 0.2 for every 7 epochs. $L_2$ regularization with $\lambda=0.001$ is used to prevent overfitting. The size of mini-batch in each epoch is set to 128. We apply early stopping as another regularization in model validation. In our practice, training usually is stopped at 15 epochs. Model parameters are initialized with uniform distribution on $(-\sqrt{u}, \sqrt{u})$ where $u$ is reciprocal of the number of weights in each layer. We implement our CNN in Pytorch \cite{paszke2017pytorch}.

To avoid model seeing images of the same subject in testing and training (since each subject generates multiple 3D fMRI images), we split the dataset at subject level. In the experiments, data are divided with stratification into training, validation and testing sets by 80\%/10\%/10\%. All images are normalized by subtracting mean image and then divided by the maximal absolute intensity value, both calculated from training data. The splitting procedure results in about 9300 images in training, 1200 in validation and testing. We repeat this procedure 10 times in experiments. To evaluate model performance, we use F1 score which is calculated as 2(precision$\times$recall)/(precision+recall) and Area under ROC curve (AUC) as the evaluation metric, and the average F1 and AUC over 10 runs are reported.

\section{Experimental Results}

\subsection{Predictive Performance}
For performance comparison, we also test several baseline classification algorithms, including random forest (RF), gradient boosting machine (GBM) and logistic regression (LR) with L2 regularization. These alternatives were tested on two types of data: (1) fetal brain functional connectivity matrices extracted by correlating resting-state fMRI time series data across 100 brain regions (Fisher z-score), (2) fetal BOLD fMRI images. For fMRI, due to the high dimensionality (dim= 39984)\footnote{We remove zero columns after flattening original 3D fMRI into a vector of size $43\times 51\times 40=87720$} and strong correlations among neighboring voxels in fMRI, we use principal component analysis (PCA) for dimension reduction before testing on baselines. We select the optimal number of PCA components as well as optimal parameters of classification models using the validation data. In our experiments, 100 components achieves good performance and including more components does not necessarily lead to performance gain.

The predictive performance (F1 score and AUC) on testing data at subject level is shown in Table \ref{tb:performance}. Note that subject level prediction is made with soft voting: probability for each subject is averaged across its corresponding 3D images. The probability threshold for classification is chosen as 0.5 for calculating F1 score. We see from Table \ref{tb:performance} that CNN trained directly on fMRI images achieves the best and robust performance (F1 0.84 with standard deviation 0.05, AUC 0.91 with 0.06), due to CNN's merit of capturing spatial information of fMRI images in classification. The improvements demonstrate that CNN can effectively capture discriminative information that could reveal associations between age effect and brain development.

\begin{table}[t]
	\centering
	\caption{Predictive performance (standard deviation) on testing subjects. Note that Fisher z-score as functional connectivity measure is calculated based on brain parcellation and not applicable for CNN. }\label{tb:performance}
	\begin{tabularx}{\textwidth}{ c *{4}{Y} }
		\toprule
		& \multicolumn{2}{c}{Fisher z-score} & \multicolumn{2}{c}{fMRI Image} \\
		\midrule
		& F1 & AUC & F1 & AUC \\
		\midrule	
		CNN & - & - & \bfseries{0.84} (0.05) & \bfseries{0.91} (0.06)\\
		\midrule 
		RF & 0.76 (0.03) & 0.69 (0.18) & 0.79 (0.06) & 0.77 (0.15) \\
		\midrule
		GBM & 0.73 (0.09) & 0.69 (0.17) & 0.81 (0.06) & 0.77 (0.18) \\
		\midrule
		LR & 0.77 (0.06) & 0.75 (0.18) & 0.77 (0.01) & 0.67 (0.14) \\
		\bottomrule
	\end{tabularx}
\end{table}

\subsection{Inference About Fetal Neurodevelopment from CNN Classification}
It was possible to discriminate older versus younger fetuses on the basis of spontaneous baseline resting-state BOLD measurements. BOLD reflects changing regional blood concentrations of oxy- and deoxy-hemoglobin, which are influenced at least in part by regional metabolic activity \cite{fukunaga2008metabolic}. Having observed that fetal age can be differentiated on the basis of these fluctuating signals suggests that patterns in baseline BOLD during resting-state reflect aspects of brain maturation. 

\begin{figure}[t]
	\centering
	\includegraphics[width=\columnwidth]{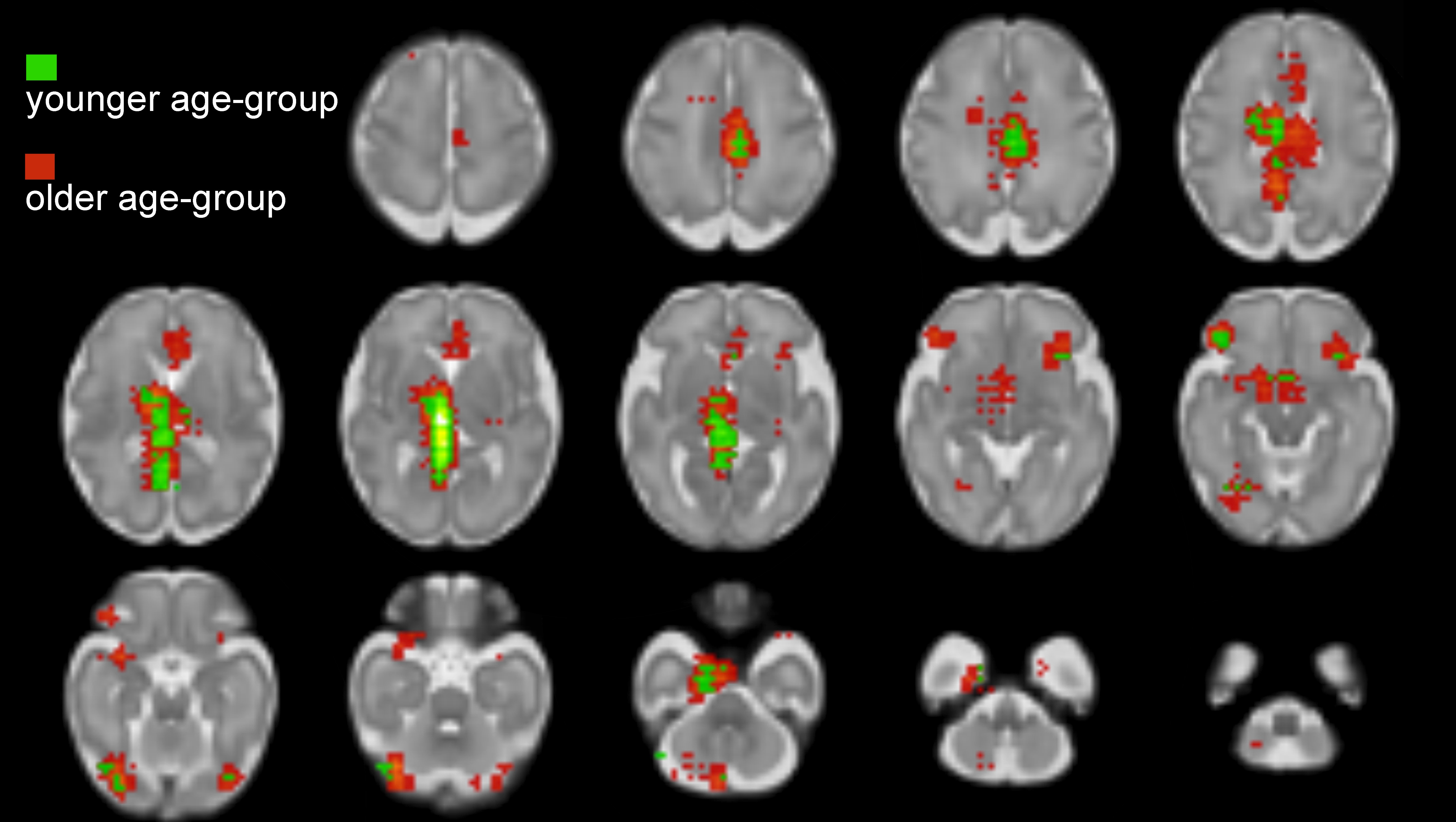}
	\caption{Sensitivity analysis for younger \textit{v.s.} older age groups. Important regions of high sensitivity scores that are associated with age effects are noted in green for younger fetuses and red for older fetuses. }
	\label{fig:senMap}
\end{figure}

Sensitivity analysis was used to identify regions within each age group where change in voxel values most altered strength of the prediction. Important regions (identified with large sensitivity scores, see Section \ref{sec:modelIntpretation}) for both younger and older fetus groups are shown in Fig. \ref{fig:senMap}. We found that sensitive regions (1) have a high degree of spatial overlap across both age groups, (2) are largely bilaterally distributed, (3) encompass brain regions that have been identified as having high baseline metabolic activity in positron emission tomography (PET) studies in human newborns and infants, including subcortex, thalamus, and medial temporal lobe \cite{chugani2018newborn}.

Specifically, across groups we observe high sensitivity in bilateral occipital cortex, bilateral ventrolateral prefrontal cortex (vlPFC), sub-cortex, posterior cingulate cortex (PCC), medial temporal lobe (MTL), thalamus, and cerebellum. In addition, baseline BOLD in anterior cingulate cortex (ACC), hypothalamus, and insula regions are important for group classification in older fetuses. 

\section{Conclusions}
In this paper, we propose a novel application of CNN for interpretation of fetal brain age effects directly from fMRI images. The predictive performance of CNN demonstrate that it can well capture associations between age and variation in BOLD signal. To better understand the relevance of our predictive CNN to fetal development, we use sensitivity analysis to isolate regions critical for CNN performance, and discovered that our most sensitive regions were regions that are high in metabolic activity in early human brain development. The experimental results reveal compelling associations and demonstrate potential promise of CNN applied to spontaneous BOLD activity as a data-driven approach to understand fetal developmental processes. 

With those identified important ROIs, in our future studies, we plan to perform seed-based analyses to examine how functional connectivities and networks of those ROIs are associated with whole-brain age effects.

\bibliographystyle{splncs04}

{\footnotesize
	\bibliography{CNNAgeRef} 
}
\end{document}